\documentclass[apjl]{emulateapj}
\usepackage{apjfonts,amssymb}
\usepackage{graphicx}
\def\apj{ApJ}
\def\apjs{ApJS}

\def\apjl{ApJL}
\def\araa{ARA\&A}

\def\aap{A\&A}

\def\prl{Phys. Rev. Lett.}
\def\pre{Phys. Rev. E}
\def\jcp{J. Comp. Phys.}
\def\lsim{~\raise0.3ex\hbox{$<$}\kern-0.75em{\lower0.65ex\hbox{$\sim$}}~}
\def\gsim{~\raise0.3ex\hbox{$>$}\kern-0.75em{\lower0.65ex\hbox{$\sim$}}~}
 
\slugcomment{Revised version \today}
\shorttitle{AMR TURBULENCE SIMULATIONS}
\shortauthors{KRITSUK, NORMAN, \& PADOAN}

\begin{document}

\title{Adaptive Mesh Refinement for Supersonic Molecular Cloud Turbulence}

\author{Alexei G. Kritsuk, Michael L. Norman, and Paolo Padoan}
\affil{Department of Physics and Center for Astrophysics and Space Sciences,
University of California, San Diego,\\
9500 Gilman Drive, La Jolla, CA 92093-0424;
akritsuk@ucsd.edu,
mnorman@ucsd.edu,
ppadoan@ucsd.edu
}

\begin{abstract}
We performed a series of three-dimensional numerical simulations of 
supersonic homogeneous Euler turbulence with adaptive mesh refinement (AMR)
and effective grid resolution up to $1024^3$ zones.
Our experiments describe nonmagnetized driven supersonic turbulent flows 
with an isothermal equation of state.
Mesh refinement on shocks and shear is implemented to cover dynamically 
important structures with the highest resolution subgrids and calibrated to
match the turbulence statistics obtained from the equivalent uniform grid 
simulations.

We found that at a level of resolution slightly below $512^3$, when a 
sufficient integral/dissipation scale separation is first achieved, 
the fraction of the box volume covered by the AMR subgrids first becomes 
smaller than unity.
At the higher AMR levels subgrids start covering smaller and smaller 
fraction of the whole volume that scale with the Reynolds number as $Re^{-1/4}$.
We demonstrate the consistency of this scaling with a hypothesis that the most
dynamically important structures in intermittent supersonic turbulence are 
strong shocks with a fractal dimension of two.
We show that turbulence statistics derived from AMR simulations and  
simulations performed on uniform grids agree surprisingly well, 
even though only a fraction of the volume is covered by AMR subgrids.
Based on these results, we discuss the signature of dissipative
structures in the statistical properties of supersonic turbulence 
and their role in overall flow dynamics.
\end{abstract}

\keywords{
ISM: structure --- 
ISM: clouds --- 
hydrodynamics --- 
turbulence --- 
methods: numerical }

\section{Introduction}
Turbulence in molecular clouds is characterized by very high integral scale
Reynolds numbers, $Re\equiv\ell_0 u_0/\nu\gsim 10^8$, 
where $\ell_0\approx10$~pc is the typical scale 
on which the turbulence is driven, 
$u_0\approx2$~km~s$^{-1}$ is the typical velocity associated with that scale,
and $\nu$ is the kinematic viscosity of the molecular gas (see Elmegreen \&
Scalo 2004 for a review).
Nonlinear interactions that dominate the dynamics of such multi-scale flows
critically depend on adequate resolution and suggest to exploit spatial and 
temporal adaptivity in numerical simulations.
While adaptive mesh refinement has previously been applied to simulate the 
evolution of individual singular structures in incompressible Euler 
turbulence \citep{pumir.90,grauer..98}, no attempts have been made so far to 
address the isotropic turbulence case with AMR.

In this letter we apply high order adaptive methods to simulate
supersonic turbulence in star forming molecular clouds with very high
resolution.
Such simulations are essential for understanding the nature of turbulent 
fragmentation that may ultimately control the stellar initial mass function 
\citep{padoan.02}.
We exploit the fact that turbulent flows are not completely chaotic.
Order is always present on both large and small scales due to the intermittent
nature of turbulence.
For instance, large under-dense voids and sharp density peaks are known to be
characteristic of supersonic turbulent flows with a ``soft'' equation of 
state \citep{passot.98}. 
We therefore expect AMR simulations of isothermal turbulence
to be beneficial in terms of computational resources.%
\footnote{Application of AMR techniques to multiphase interstellar 
turbulence simulations, where the effective adiabatic 
index determined by the balance between heating and cooling at high densities
falls below unity, appears to be even more promising.}
We show that in fact AMR technique can be profitably applied to
turbulence simulations in contrast with the prevailing {\em belief} 
that in such studies the adaptive mesh does not help as the fine 
structures emerge through the entire computational domain.

\begin{figure*}
\epsscale{1.1}
\centerline{\plottwo{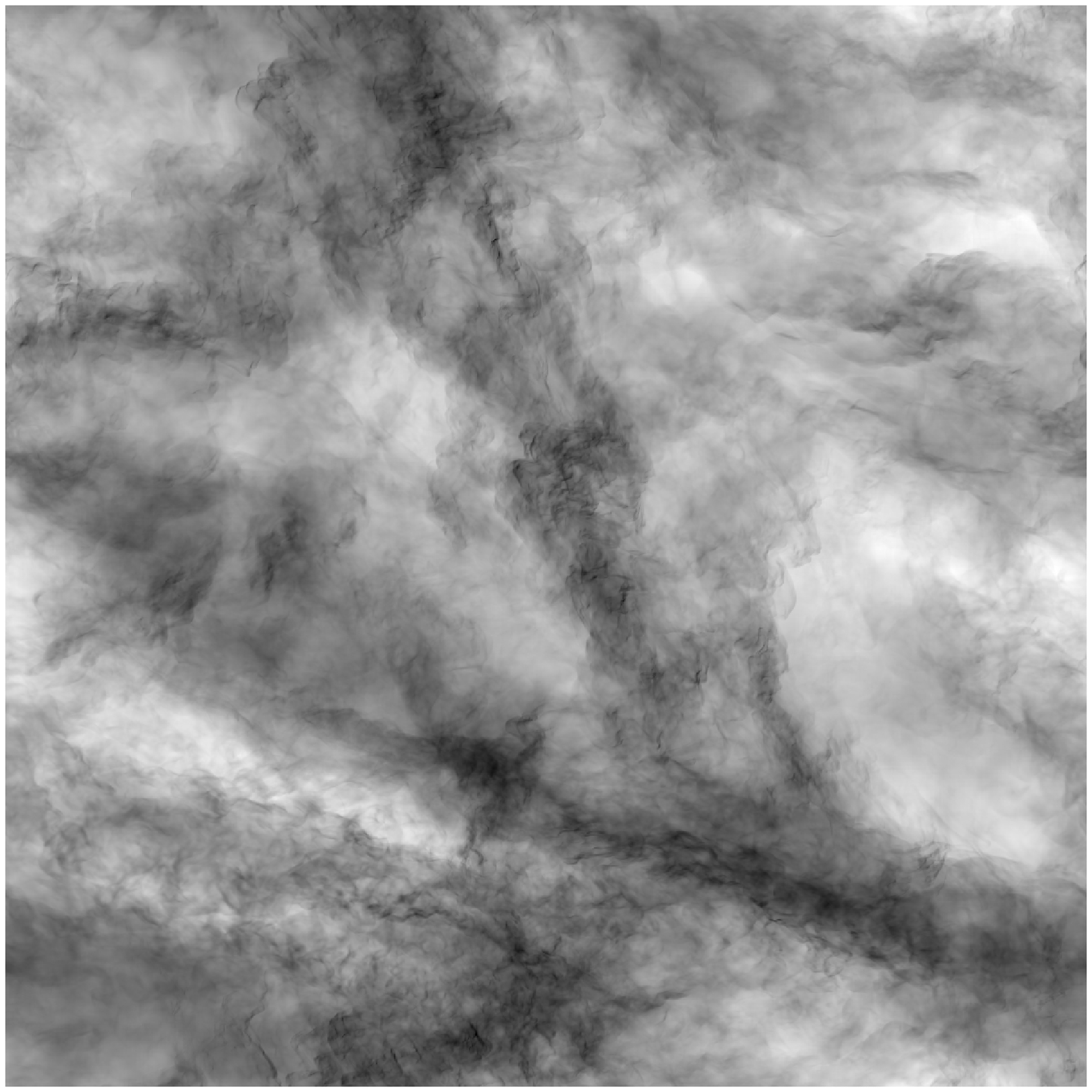}{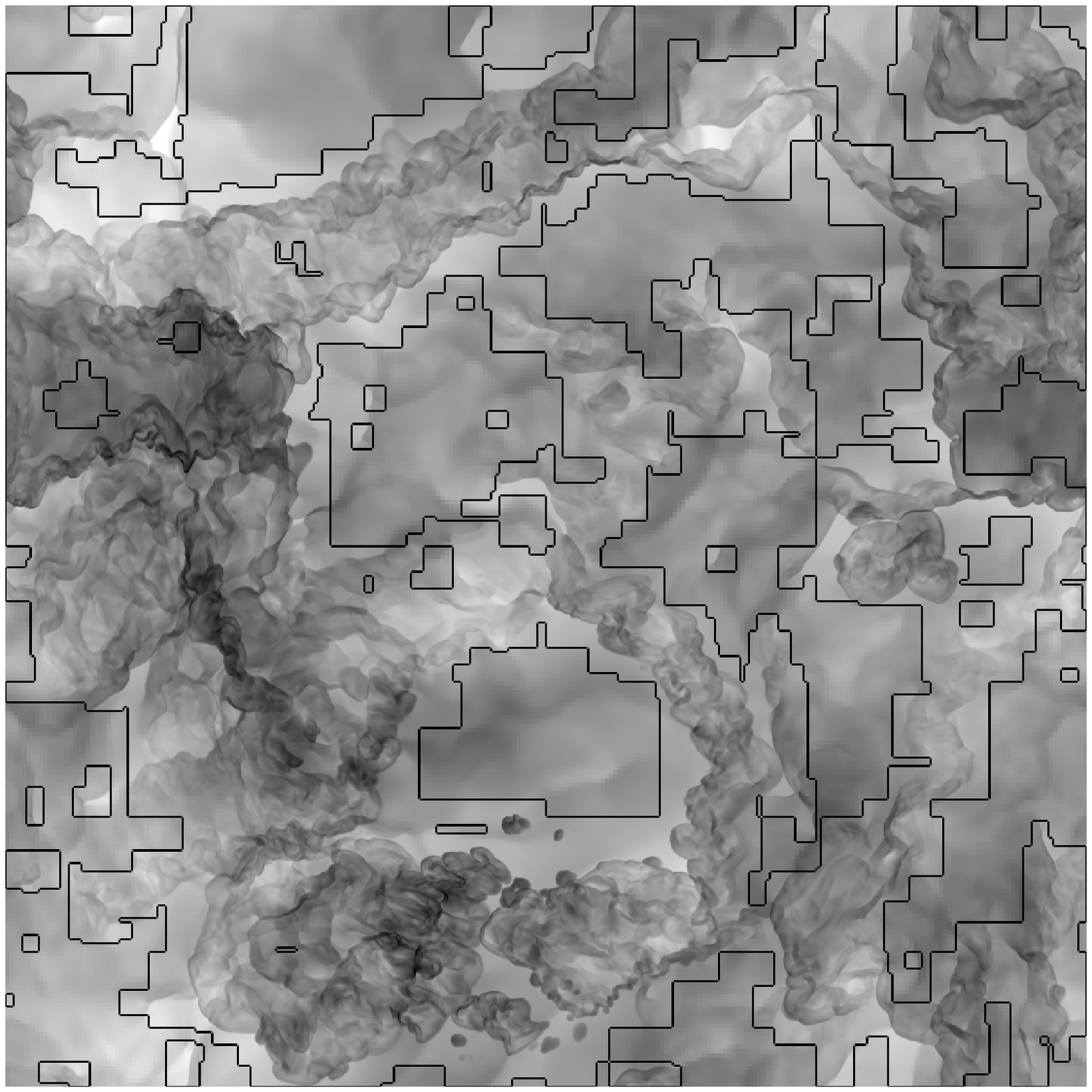}}
\caption{Gas density from AMR simulation of supersonic Mach 6 turbulence 
with effective resolution of $1024^3$ at time $t=6\;t_{dyn}$.
{\em Left panel:} 
a projection of the density field
through the computational box (log scale, dense regions are dark).
{\em Right panel:}
a thin slice from the same density 
distribution with contours showing the patches of the base grid {\em not 
covered} by the AMR subgrids.
The total volume covering factor of the first level AMR subgrids is 
approximately 65\%.}
\label{plates}
\end{figure*}
 
\section{Turbulence, intermittency, and adaptive meshes}

It follows from the \citet{kolmogorov41} phenomenology for turbulent cascade 
(K41) that the inertial range spans an interval of scales 
$\mathcal{N}\equiv\ell_0/\eta\sim\left[\nu^3/(\ell_0^3 
u_0^3)\right]^{-1/4}\sim Re^{3/4}$, 
where $\eta\sim(\nu^3/\varepsilon)^{1/4}$ is the Kolmogorov dissipation scale,
and the energy cascade rate $\varepsilon\sim u_0^3/\ell_0$.
Hence, the number of zones per integral scale $\ell_0^3$ required to 
model a three-dimensional turbulent flow using a finite-difference 
method is $\mathcal{N}^3\propto Re^{9/4}$.
The storage resource for a numerical experiment of this sort would 
also grow as $\mathcal{S}\propto Re^{9/4}$, 
while the number of CPU hours needed to simulate 
high-Reynolds-number flows on a uniform grid for a given number 
of large eddy turnover times would scale as $\mathcal{T}\propto Re^3$ 
(e.g. Frisch 1995).

However, the above calculation implicitly assumes that the inertial
range flow is completely chaotic and does not contain any coherent
structures, which is often not the case in experimental high-$Re$ flows.
In fact, both laboratory experiments and numerical simulations indicate the
presence of some order in the flows on both small $\ell\sim\eta$ and large 
$\ell\sim\ell_0$ scales.
Since turbulence is intermittent, the K41 theory gives only 
an approximation to its statistical properties and requires corrections to 
reproduce experimental measurements of high-order statistics.
One way to add a form of intermittency to the K41 model is known as 
the $\beta$-model \citep{frisch..78}.
It assumes that the fraction of the space occupied by the
`active' eddies of size $\ell$ scales as $(\ell/\ell_0)^{3-D}$, where
$D$ is interpreted as the fractal dimension of small-scale dissipative 
structure.
In the framework of the $\beta$-model the dissipation scale is a function of 
$D$, therefore $\mathcal{N}_{D}\equiv\ell_0/\eta(D)\sim Re^{3/(1+D)}$.
One can easily recover the familiar K41 result from this formula by setting 
$D=3$ which corresponds to the zero-intermittency limit.
Thus, the scaling for the storage depends on the fractal dimension of the 
small-scale dissipative structure, 
$\mathcal{S}_D\propto Re^{3D/(1+D)}$.

There are good reasons to assume $D\approx1$ in the incompressible case, 
where vortex filaments are the most singular dissipative structures
\citep{she.94,dubrulle94}, and $D\approx2$ for supersonic compressible 
turbulence, where the dissipative structures are shocks \citep{boldyrev02}.
For these fractal dimensions,
$\mathcal{S}_{1}/\mathcal{S}\sim Re^{-3/4}$ and 
$\mathcal{S}_{2}/\mathcal{S}\sim Re^{-1/4}$, so the advantage of adaptive
methods for simulations of high-$Re$ turbulent flows is quite clear.

There is a simple direct analogy between the process of building an AMR 
hierarchy 
to resolve the turbulent structures of the fractal dimension $D$ and the 
box-counting method for determination of that same fractal dimension.
The fractal dimension is given by the relation 
$n(\delta)\propto \delta^{-D}$, where $n(\delta)$ is the number of boxes of 
linear size $\delta$ needed to cover the set of singular structures.
One can think of the box size $\delta$ as a function of the level 
number $l$ in the AMR hierarchy, $\delta_l=\delta_0 f^{-l}$, 
where $\delta_0$ is the zone size on the base grid and $f$ is the refinement 
factor (usually $f=2$ or 4).
Then the number of zones needed to cover the structures on level $l$ would grow
as $n_l\propto\delta_l^{-D}\propto f^{lD}$, and the volume covering factor
of level $l$ subgrids would scale with the level number as 
$F_l\propto f^{l(D-3)}$.
Assuming $D=2$, one gets $F_l\propto f^{-l}$.
Having chosen an appropriate resolution for the base grid and $f=4$, 
one can predict the covering factors for $l=1$, 2, and 3-subgrids 
to be 0.25, 0.06, and 0.015, respectively.
For incompressible turbulence, the same factors are valid if $f=2$.

At what base grid resolution AMR first becomes beneficial for turbulence
simulations?
As with the box-counting method, the expected scaling can be attained only with
a high enough resolution.
It is necessary to provide integral/dissipation scale separation for the 
inertial range to become extended in the wavenumber space.
In simulations of compressible Euler turbulence with PPM this can be achieved
at resolutions somewhere in between $256^3$ and $512^3$ (schemes with
higher numerical diffusivity than PPM would require many more zones), while 
for compressible Navier-Stokes turbulence one needs about $4^3$ times higher 
resolution since a larger range of scales is needed to account for the 
physical dissipation \citep{sytine....00}.
Therefore, in PPM simulations of Euler turbulence the advantage of AMR 
comes into play earlier than in those of Navier-Stokes turbulence.
On the other hand, due to the difference in scaling behavior,
AMR is expected to save more resources in numerical experiments with 
incompressible flows at very high Reynolds numbers.
In the following sections we show that in practice for Mach 6 isothermal 
turbulence the advantage of AMR is indeed first noticeable at a resolution 
of about $512^3$, and the scaling relations derived above are consistent with
our numerical experiments which involve the {\sl Enzo} 
code \citep[and references therein]{oshea......04}.

\section{Numerical approach}
The {\sl Enzo} code uses a direct Eulerian formulation of
the Piecewise Parabolic Method \citep[PPM]{colella.84}
to solve the equations of gas dynamics on a hierarchy of structured 
adaptive meshes \citep{berger.89}.

In order to mimic the conditions in molecular clouds, we adopt a 
quasi-isothermal equation of state with the ratio of specific heats 
$\gamma=1.001$. 
To test our PPM implementation in this highly compressible low-$\gamma$ regime
and at high Mach numbers, we ran a number of tests with and without AMR, 
including simple shock tube tests described in \citep{balsara94}.
We found very good agreement with the exact solutions and with results 
obtained with a Riemann solver for purely isothermal gas dynamics. 
The temperature variations in tests with Mach 6 flows, due to the
small deviation of $\gamma$ from unity, were typically below 0.1\%. 

\begin{figure}
\epsscale{1.15}
\centerline{\plotone{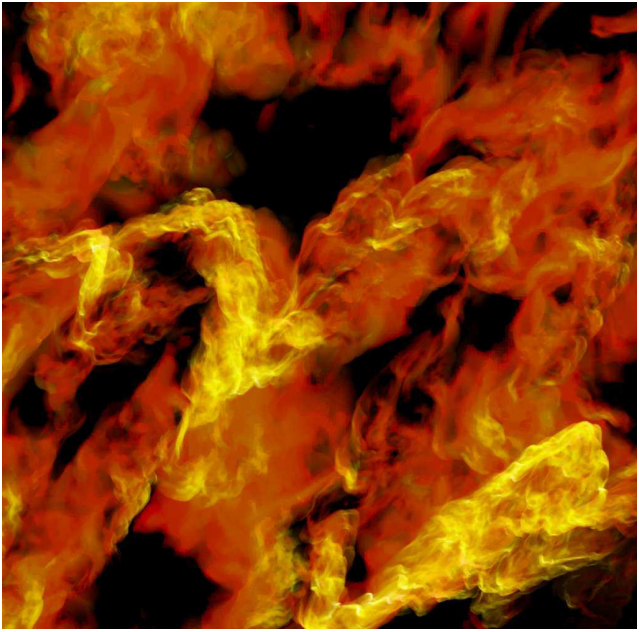}}
\caption{Volumetric rendering of the gas density in a 100-zone-thick slice
through a subsample of the computational domain.
V- and U-shaped shocklets or ``Mach cones'' are 
the most common structures in supersonic isothermal hydrodynamic turbulence. 
As in a hierarchy of vortices in incompressible turbulence, large-scale 
Mach cones are broken into smaller and smaller ones down to the finest
resolved scale. Mach angles of the largest cones visible in this snapshot 
correspond to relative fluid velocities of $(2-3)\,c_s$. }
\label{bow}
\end{figure}
 
\begin{figure}
\epsscale{1.16}
\vspace{-6.4cm}
\centerline{\plotone{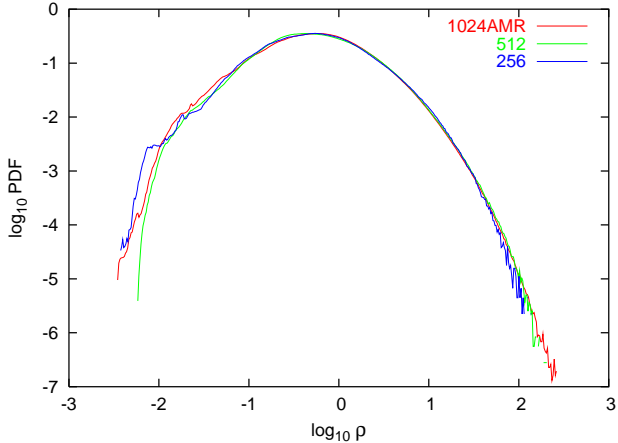}}
\caption{Comparison for turbulence statistics based on $256^3$ and $512^3$
uniform grid simulations and an AMR simulation with effective grid resolution
of $1024^3$ at time $t=6\;t_{dyn}$. 
Probability density functions (PDFs) for the gas 
density.}
\label{dpdf}
\end{figure}
 
\begin{figure}
\epsscale{1.16}
\vspace{-6.4cm}
\centerline{\plotone{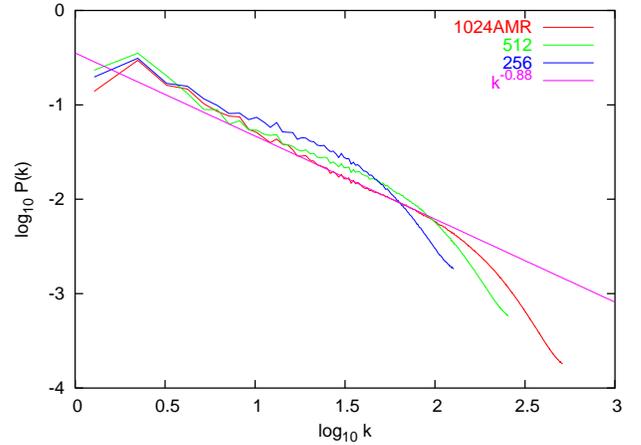}}
\caption{Same as in Fig. \ref{dpdf}, but for
the gas density power spectra.}
\label{dpow}
\end{figure}
 
The AMR subgrids are placed where needed to resolve shocks with large
pressure jumps and regions of strong shear.
We identify shocks using the PPM shock detection algorithm. 
In addition, we also use a norm of the velocity gradient matrix 
$\parallel \partial_i u_j\parallel$ to account for shear.
This matrix norm is similar to the Frobenius norm, except that it does not
include the contribution from diagonal elements.
To get a better AMR efficiency for simulations of turbulent flows (which 
involve a large number of subgrids) we use mesh refinement by a 
factor of 4.
We obtained a very good agreement between our nonadaptive and AMR runs for 
refinement on shocks with pressure jumps $\Delta p/p\geq2$.
When a combination of shocks and shear controlled the refinement, a minimum 
pressure jump of 3 was sufficient, while the application of the second 
refinement criterion accounted for some $20$\% more zones to be flagged.
Note that these values are given only for orientation, while the actual lower 
bound for pressure jumps in refined shocks would depend on the AMR 
implementation.

To maintain the turbulent kinetic energy in the computational box at a given
level, we use large-scale solenoidal force per unit mass with a fixed spatial 
pattern and a constant power in the range of wavenumbers $k\in[1,2]$.

\begin{figure}
\epsscale{1.16}
\vspace{-6.4cm}
\centerline{\plotone{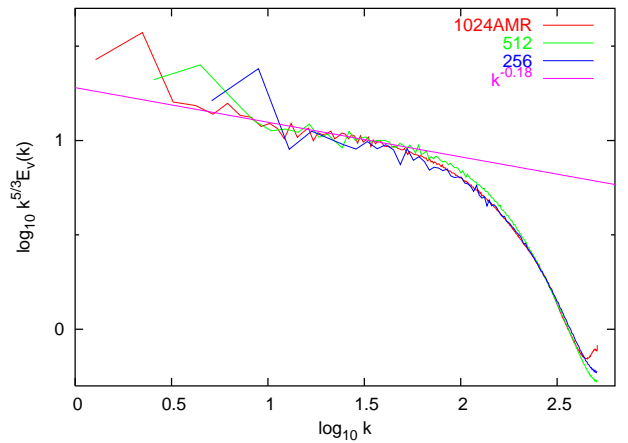}}
\caption{Same as in Fig. \ref{dpdf},
but for the compensated total velocity power spectra,
scaled so that the curves match in the numerical
dissipation range of PPM where they are invariant.}
\label{vpow}
\end{figure}
 
\section{Turbulent structures at high resolution}
We first performed a series of nonadaptive simulations of driven
turbulence with an rms Mach number of 6 varying 
the uniform grid resolution from $128^3$ up to $512^3$ zones.
We started the simulations with a uniform gas density and ran them for 
6 dynamical times\footnote{$t_{dyn}\equiv L/(2M)$, where $L$ is the box size,
$M$ is the rms Mach number, and the sound speed $c_s$ is unity.} to follow the
development and saturation of the turbulent flow.
We then restarted the $128^3$ run as an AMR simulation with the base grid 
of $128^3$ and one level of refinement by a factor of 4,
ran it from 4.8 to 6 $t_{dyn}$, and compared the results with our
uniform grid run with the equivalent resolution of $512^3$ zones.
After a few iterations we found that refinement on shocks with pressure jumps
above $\approx2$ gives us velocity power spectra consistent with the spectra 
from nonadaptive run with the same effective resolution.
The velocity statistics appear to be more sensitive to the refinement
criterion then the density PDF or the density power spectrum.
Since the volume covering factor of the first level subgrids in this AMR 
simulation was about 90\%, it remained unclear whether it is the high covering
fraction that provides convergence or it is indeed the right refinement 
criterion.
Before switching to higher resolution base grids (it is wasteful to run
AMR covering 90\% of the volume) we restarted the same AMR run from
$t=6\;t_{dyn}$ allowing for two levels and got an estimate of the covering
fraction for the second level about 34\% at effective resolution of 
$2048^3$.
This number should be considered as a lower limit since it will 
grow by a few percent over the following $t_{dyn}$, while the flow gets 
fully resolved on the second level.

We then repeated the same experiment with the $256^3$ base grid and one level 
of refinement, with however a slightly different refinement recipe which
also included shear, see Section 3 for details.
With effective resolution of $1024^3$ this AMR run is the largest
simulation to date of supersonic turbulence in molecular clouds.
Fig. \ref{plates} shows a snapshot of the gas density field from this run.
From the left panel, where we show a distribution of the column density, 
it may indeed look like the fine structures emerge through the entire 
computational domain.
However, the morphology of the projected density field is quite different
from what one can see in a slice, highlighting a significant loss of 
information built into the projection procedure.
As one can see from a thin slice through the box shown in the 
right panel, AMR does help in this case since turbulence is very intermittent.
The first level subgrids do not cover the under-dense voids and also some 
smaller windows in the high and intermediate density gas where the strong 
shocks are absent.
These regions do not contribute much to cascading the energy from large to 
small scales as their share in the turbulence 
statistics is the same whether they are refined or not.
Strong shock interactions and associated nonlinear instabilities create a
very sophisticated multiscale pattern in the dynamically active regions 
(Figs. \ref{plates}, \ref{bow}) that is morphologically similar to what
is observed in molecular clouds.
This pattern is missing in numerical simulations at lower $Re$. 
We identify Mach cones and U-shaped shocklets as self-similar elements of 
this pattern (Fig. \ref{bow}).
These effects of nonlinear shear instabilities resolved in our simulations
can explain the lack of large-scale shock signatures in the observations 
of molecular gas by \citet{brunt03}.

While we do not have an equivalent nonadaptive simulation, 
the comparison with PDFs and power spectra from $256^3$ and $512^3$ runs 
shown in Figs. \ref{dpdf} -- \ref{vpow} speaks for itself.
The density PDFs can be very precisely fitted by a log-normal distribution 
and the AMR data match those from the two nonadaptive runs, although the 
superior resolution of the AMR run provides better sampling for the high 
end of the density distribution.
Interactions of strong counter-propagating shocks are responsible for
intermittent oscillations in the high density wing of the PDF, which are
well resolved in our AMR simulation. 
These interactions also cause transient strong rarefactions lagging behind
in time.
As a result, on time-scles short compared to the dynamical time, the density 
PDF slightly wanders around its average log-normal representation at the 
highest densities and displays large temporal oscillations in its low-density 
end, see  Fig. \ref{dpdf}.
The same processes reveals itself in correlated variations of the 
three-dimensional power spectrum of density in the inertial range.
The slope lies somewhere between $-0.8$ and $-0.9$ in our nonmagnetized 
models, see Fig. \ref{dpow}.
The spectrum gets shallower upon the collisions of strong shocks, when the
PDF's high density wing rises above the average log-normal representation.

\lastpagefootnotes

The density power spectrum builds up quickly at high wavenumbers,
after we switch the AMR machinery on, simply because PPM starts resolving 
the shocks better.
However, the relaxation of the velocity power proceeds slower since
it takes about one dynamical time for the resolved local sharp density 
structures to get through nonlocal dynamical interactions involving multiple 
scales.
Only then the inertial range really extends to smaller scales.
When AMR is first activated, the velocity power is insufficient at $k\gsim25$
and scales approximately as $k^{-2}$ in this range.
It then steadily accumulates at those frequencies for about $t_{dyn}$ and 
saturates exactly at the level predicted by our nonadaptive 
simulations, see Fig. \ref{vpow}.
The slope of the velocity power spectrum for the snapshot shown is
about $-1.85$, i.e. somewhat steeper than a slope of $-1.74$ predicted by 
\citet{boldyrev02} for supersonic turbulence assuming $D=2$ and assuming the 
third order structure function exponent equal to unity, as in incompressible 
turbulence.\footnote{%
Averaging over independent realizations of the turbulent flow is 
needed to obtain reliable estimates for the scaling exponents.
This lies outside the scope of this letter which is primarily focused
on the applicability of adaptive methods.}
According to the formalism developed by \citet{dubrulle94}, a slope of $-1.85$
would imply $D\approx2.3$.
This is consistent with the dimensionality
of dissipative structures that can be independently estimated using the 
volume covering fractions of AMR subgrids at different levels of resolution 
(90, 65, and 34\% at $512^3$, $1024^3$, and $2048^3$, respectively). 
Such estimate returns the same value of $D\approx2.3$.
Within the uncertainties, both independent estimates agree with the 
observationally determined fractal dimension of molecular clouds 
$D=2.3\pm0.3$ \citep{elmegreen.96}.

\section{Conclusions}

While details of AMR implementation may vary and may have to be further 
refined to reproduce higher order statistics, it is clear that 
adaptivity in both space and time is indispensable for numerical experiments 
with homogeneous isotropic supersonic turbulence at very high Reynolds numbers 
with grid resolutions $\gsim 1024^3$.

Our simulations reveal a pattern of small-scale structures that is completely
missing at lower $Re$.
These structures originate in nonlinear instabilities inherent in isothermal 
supersonic flows and may control the scaling properties of turbulence and
the fractal dimension of the dynamically important structures.
Since the presence of magnetic fields can modify the unstable modes, it
cannot be taken for granted that magnetized turbulence should have the same
scaling properties.

Numerical experiments at very high Reynolds numbers are crucial for studies
of turbulence in molecular clouds.

\acknowledgments

This work was partially supported by NRAC allocation MCA098020S
and utilized computing resources provided by the San Diego Supercomputer
Center.

\end{document}